\documentclass[conference]{IEEEtran}
\IEEEoverridecommandlockouts
\usepackage{cite}
\usepackage{amsmath,amssymb,amsfonts}
\usepackage{algorithmic}
\usepackage{graphicx}
\usepackage{textcomp}
\usepackage{xcolor}
\def\BibTeX{{\rm B\kern-.05em{\sc i\kern-.025em b}\kern-.08em
    T\kern-.1667em\lower.7ex\hbox{E}\kern-.125emX}}

\begin{document}
\title{SDN - Architectural Enabler for Reliable Communication over Millimeter-Wave 5G Networks}

\author{\IEEEauthorblockN{Biswa P. S. Sahoo, Chung-Wei Weng, and Hung-Yu Wei\IEEEauthorrefmark{1}}
\IEEEauthorblockA{Graduate Institute of Electrical Engineering, National Taiwan University, Taipei, Taiwan\\
\IEEEauthorrefmark{1}Email: hywei@cc.ee.ntu.edu.tw}}

\maketitle

\begin{abstract}
Millimeter-wave (mmWave) frequency bands offer a new frontier for next-generation wireless networks, popularly known as 5G, to enable multi-gigabit communication; however, the availability and reliability of mmWave signals are significantly limited due to its unfavorable propagation characteristics. Thus, mmWave networks rely on directional narrow-beam transmissions to overcome severe path-loss. To mitigate the impact of transmission-reception directionality and provide uninterrupted network services, ensuring the availability of mmWave transmission links is important. In this paper, we proposed a new flexible network architecture to provide efficient resource coordination among serving basestations during user mobility. The key idea of this holistic architecture is to exploit the software-defined networking (SDN) technology with mmWave communication to provide a flexible and resilient network architecture. Besides, this paper presents an efficient and seamless uncoordinated network operation to support reliable communication in highly-dynamic environments characterized by high density and mobility of wireless devices. To warrant high-reliability and guard against the potential radio link failure, we introduce a new transmission framework to ensure that there is at least one basestation is connected to the UE at all times. We validate the proposed transmission scheme through simulations.
\end{abstract}



\section{Introduction}\label{sec:intro}
The next-generation wireless communications system, popularly known as fifth generation (5G), is envisioned to use frequencies in the range of 30$-$300 GHz band, which often referred to as millimeter-wave (mmWave) bands. Recently, mmWave frequency bands have raised increasing attention to support emerging wireless broadband-access and cellular system evolution with its achievable coverage comparable to traditional systems~\cite{Pi2011, WengTVT01, WengTVT02, Sahoo2017}. A vast amount of underutilized spectrum in this frequency range targeting multi-gigabit mobile broadband is the first promise mmWave technology has to offer. Different from the wireless systems using lower frequency bands, the mmWave transmission is featured with the short wavelength, larger transmission bandwidth. However, recent studies~\cite{Akdeniz2014, Shokri-Ghadikolaei2015, Weng2018} show that the mmWave systems suffer from several limitations such as sensitive to blockage against most solid materials, and short transmission range. These challenges were partly addressed by recent publications.
\par
The International Telecommunication Union (ITU) has classified 5G mobile network services into three categories, viz., eMBB (enhanced Mobile Broadband), mMTC (massive Machine-Type Communications), and uRLLC (ultra-Reliable and Low-Latency Communications)~\cite{IMTVision,r2-164130}. However, the harsh propagation environment at such high frequencies and unreliable mmWave transmission link make it hard to provide reliable services. Additionally, the user mobility disrupts gains of mmWave which may impact both reliability and latency requirements. Thus, providing ultra-high reliable services in high-mobility scenarios requires overcoming several challenges in fast-varying networks. First, the frequent hand-off between base stations (BSs) in ultra-dense deployments which introduces additional control signaling and latency overhead. Second, the overall end-to-end quality of experience which requires ultra-reliability, low packet loss, and a stable data rate at the end user. In addition, it is essential to guard against potential radio link failures and network infrastructure failures.
\par
To address the aforementioned issues related to high mobility scenarios and ultra-dense deployments in mmWave communication, joint transmission scheme such as coordinated multi-point (CoMP) \cite{Lee2012}, might serve as an enabling solution. However, the CoMP scheme requires a close coordination between BSs, which may not feasible all the time. Additionally, to facilitate the fast switching of the cells, a highly detailed feedback is required on the channel properties at the user equipment (UEs). Alternatively, multi-connectivity \cite{Drago2018} can also be used to satisfy the reliability requirements by ensuring multiple alternative serving paths to a UE. Multi-connectivity uses different radio access technologies, such as LTE at sub-6 GHz frequencies and New Radio (NR) at mmWave frequencies, to provide an alternative path when one of the links is not available. However, in the current cellular architectures, such coordination of BSs is very limited~\cite{Lin2017}. Additionally, applying these joint transmission or coordination based schemes over mmWave frequencies band may not be suitable, as it introduces additional control signaling which may limit the overall system performance. 
\par
Although Software-Defined Networking (SDN) technology is the recent trend which can overcome the pitfalls; however, their applicability in next-generation wireless networks are still to be fully explored and exploited \cite{Sahoo2018}. SDN provides numerous benefits with the provisioning of programmability to reduced capital and operational expenses for wired networks~\cite{Kar2016, Kar2018}. The SDN offer an essential platform to solve the problems of closed, inflexible, and difficult to model and standardize a network architecture~\cite{Chaudhary2018}. The objectives of SDN is to provide flexible network management, control, and high resource utilization and adapts the diverse requirement of 5G wireless systems. Thus, SDN can act as an architectural enabler for the 5G system for efficient coordination between BSs and UE~\cite{Akyildiz2015}.
\par
In this paper, we aim is to provide reliable communications in a high-mobility scenario and ultra-dense deployments. In such scenarios, frequent handover interruption time and fast-varying mmWave link impact both the reliability and latency requirements. Although existing techniques, such as Hybrid automatic repeat request (hybrid ARQ or HARQ)~\cite{Woltering2014}, can provide a certain degree of reliability, this may violate the latency requirements. To satisfy high-reliability and guard against the potential radio link failure, we introduce SDN as the architectural enabler for efficient coordination among multiple BSs and UE to meet the service requirements. Moreover, we adopted the features of network function virtualization to estimate the fast-varying channel conditions in mmWave networks. Our contributions, in this paper, can be summarized as follows:

\begin{itemize}
\item We designed an SDN-enabled architectural framework to provide a seamless mobility in the mmWave 5G scenario. 
\item We introduce a framework to form a serving cluster for UE by extending the SDN functionality in mmWave 5G networks to reduce the handover interruption time.
\item We introduce a joint transmission scheme to ensure reliability during UE mobility.
\item We evaluate the system performance by performing simulations in high mobility scenario.
\end{itemize}

The paper is organized as follows. Section~\ref{sec:related} provides an overview of the state of the art. Section~\ref{sec:system} introduces the system model and provides the proposed architecture for reliable mmWave communications. Section~\ref{sec:evaluation} presents the performance evaluation. Finally, Section~\ref{sec:con} concludes the paper.

\section{Related Work}\label{sec:related}
In literature, the software-defined architectures are well-studied in both wired and wireless networks. Recently, few SDN architectures \cite{Lin2017, Akyildiz2015, Chaudhary2018, Khan2018} are explored in wireless networks.  For achieving reliable and low latency communication in 5G mmWave cellular networks, we need to revisit the legacy cellular network design from its core itself. The end-to-end (E2E) services in cellular networks suffer delay primarily from the packet routing via the core network. To reduce this E2E delay, the core network function needs to be moved closer to the edge devices \cite{Ku2017}.
\par
The authors in \cite{Akyildiz2015} proposed a Wireless-SDN solution called SoftAir for 5G systems which enables an efficient coordination among the remote radio heads (RRHs) via software-defined architecture. Using the architecture proposed in~\cite{Akyildiz2015}, the authors in \cite{Lin2017} introduced a dynamic BS formation to solve the none-line-of-sight (NLoS) coverage problems in 5G mmWave communication. The authors jointly optimize RRH-UE associations and beamforming weights of RRHs to maximize the UE sum-rate while guaranteeing QoS and system-level constraints. However, both these works~\cite{Akyildiz2015, Lin2017} are limited given the reliable communications over mmWave networks. The authors in \cite{Yao2017} proposed an integrated framework is to reduce the outage probability while maintaining seamless service continuity by serving UE with multiple-beams simultaneously. While the proposed approach in~\cite{Yao2017}, can provide a certain degree of improvement in throughput, but beamforming association with UE is challenging. Although the handover problem has been studied widely for wireless communication; however, the cell switching is made with control signaling exchange among several coordinated entities. The other requirement is for very close coordination between several entities (BSs) to facilitate the reliable connection and seamless transition between cells.
\par
To the best of our knowledge, the software-defined architecture at mmWaves for reliable transmission and latency constraint communication has not been studied yet. CoMP and multi-connectivity were studied to satisfy the quality of service constraints, but without enabling the true capabilities of SDN and at high frequencies. On the other hand, a joint transmission scheme is to be used to ensure reliability during UE mobility.

\section{System Overview}\label{sec:system}
In this section, we first introduce the proposed SDN-enabled mmWave 5G network architecture. Second, we describe the main propagation characteristics of the mmWave bands used in the next sections of the paper.

\subsection{SDN-Enabled mmWave 5G Network}\label{subsec:w-sdn}
\begin{figure}[htbp]
\centerline{\includegraphics[width=0.95\columnwidth]{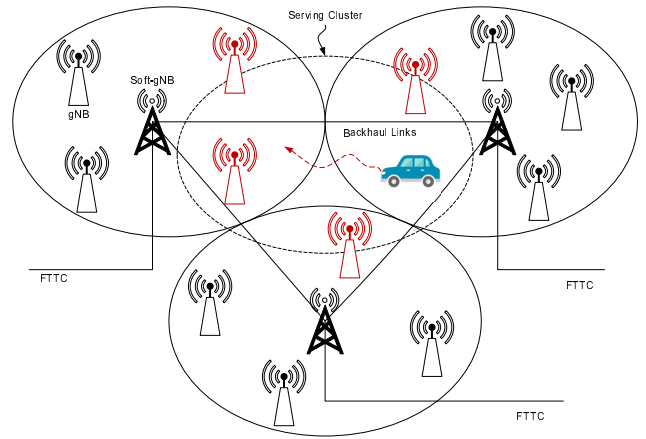}}
\caption{SDN-enabled architecture for 5G mmWave networks.}
\label{fig:network}
\end{figure}

We consider a small-cell scenario supporting multi-user, multi-cell downlink operation as shown in Figure \ref{fig:network}.  All next generation NodeBs (gNBs) are assumed to be uniformly geo-distributed. The gNB located in the cell-center is assumed to be SDN-enabled and we call it Soft-gNB. All gNBs are equipped with highly directional antennas, which are remotely controlled by Soft-gNBs and serve downlink transmissions to UEs. The backhaul traffic of a gNB is relayed by adjacent gNBs over mmWave links. All backhaul traffic from adjacent gNBs will be cooperatively forwarded to the Soft-gNB which is connected to the core network by fiber to the cell (FTTC) links.

The networks operates in slotted time $t \in \{0, 1, 2, \dots\}$, where $\mathcal{B} \triangleq \{1, \dots, B\}$ denotes the set of gNB, $\mathcal{U} \triangleq \{1, \dots, U\}$ denotes the set of UEs, and $\mathcal{L} \triangleq  \{1, \dots, L\}$ denotes the set of transmission links from gNB to each UE, indexed by $b$, $u$, and $\ell$ respectively. We denote the cardinalities of these sets as $B$, $U$, and $L$ respectively.
\par
As shown in Figure~\ref{fig:network}, a UE moving with varying speed through the edge of multiple cells.  The serving set of gNBs and the non-serving gNBs are colored 'blue' and 'black', respectively.  In this example, the serving cluster spans three cells.  As the UE moves across the cells, set of serving gNBs moves with the UE. In this approach, the UE is always a ``cell center" UE. Thus, out proposed architecture can ensure that there is at least one gNB connected to the UE at all times with strong received power to produce higher performance.
\par
Even though only one gNB may be sufficient to ensure successful transmission, multiple gNBs should always be used for uRLLC.  The multiple gNBs can form a serving cluster for the UE.  During UE mobility, the serving set of gNBs is continuously updated.  As the UE approaches a non-serving gNB, the gNB can be added to the serving cluster before the link is sufficiently strong enough for reliable transmission.  Conversely, as the UE moves out of coverage of serving gNBs, the gNBs can be removed from the serving cluster.  Since the UE may be crossing multiple hyper-cells, the serving cluster of gNBs should be able to span multiple hyper-cells.
\par
Suppose each UE is served by a serving cluster, and a gNB can serve multiple UEs simultaneously. We introduce the following binary variables as the indicators to express the association between gNBs and UEs.  In particular, while gNBs can be active to serve UEs or idle, $\{\alpha_b, b \in \mathcal{B}\}$ denotes the activity of gNBs as \cite{Lin2017}

\begin{equation}
\alpha_b = 
\begin{cases}
    1,        & \text{the } b\text{th gNB is in Serving Cluster}\\
   0,        & \text{otherwise}
\end{cases}
\end{equation}

Also, $\{\beta_{ub}, u \in \mathcal{U}, b \in \mathcal{B}\}$ denotes the association between gNBs and UEs as

\begin{equation}
\beta_{ub}= 
\begin{cases}
    1,        & \text{the } b\text{th gNB is in Serving Cluster}\\
    0,        & \text{otherwise}
\end{cases}
\end{equation}

Furthermore, to characterize the serving cluster of gNBs, the clustering indicators $\{N_{ub}, u \in \mathcal{U}, b \in \mathcal{B}\}$ are introduced as

\begin{equation}
N_{ub}= 
\begin{cases}
    1,        & (u, b) \in \mathcal{L}\\
    0,    & (u, b) \notin \mathcal{L}
\end{cases}
\end{equation}

where $\mathcal{L} = \{(u, b) | u \in \mathcal{U}, b \in \mathcal{B}_u\}$ denotes the predetermined set of feasible association and $\mathcal{N}_u$ denotes the set of near gNBs for the $u$th UE, which can be determined based on the distance or channel gain from gNBs to each UE. From these variable definitions, we can obtain the equality $\alpha_b = 1 - \Pi_{u=1}^{I}(1-\beta_{ub}N_{ub}), \forall b \in \mathcal{B}$ and two sets of association constraints between gNBs and UEs as follows:

\begin{equation}
\alpha_b \geq \beta_{ub} N_{ub}, \forall u \in \mathcal{U}, b \in \mathcal{B};
\end{equation}

\begin{equation}
\sum_{b=1}^{B} \beta_{ub} N_{ub} \geq 1, \forall u \in \mathcal{U}
\end{equation}

Eq. (4) implies that a gNB is in an active mode if it is associated with at least one UE. Eq. (5) ensures that each UE is served by at least one gNB.

\subsection{Millimeter-Wave Channel Models}\label{subsec:chmod}
In the introduced mmWave communication system, we adopt a codebook-based beamforming at that used for downlink transmission from gNBs to UEs. Let the time-varying channel gain between a UE-gNB pair be represented by a matrix $\mathbf{H}$.  For the small-scale fading model, each of gNB $B$ can then be synthesized with a reasonable number, say L = 2, of subpaths.  Each subpath will have horizontal and vertical AoAs, $\theta_{bl}^{rx}, \phi_{bl}^{rx}$,  and horizontal and vertical AoDs, $\theta_{bl}^{tx}, \phi_{bl}^{tx}$, where $b = 1, \dots, B$ is the gNB index, $l = 1, \dots, L$ is the subpath index of gNB to UE within the cluster. Assuming, there are $n_{rx}$ RX antennas and $n_{tx}$ TX antennas, the channel gain matrix as \cite{Akdeniz2014}

\begin{equation}
\mathbf{H}(t) = \frac{1}{\sqrt{L}} \sum_{b=1}^{B} \sum_{l=1}^{L} g_{bl}(t)~\mathbf{u}_{rx}\Big(\theta_{bl}^{rx}, \phi_{bl}^{rx}\Big)~\mathbf{u}_{tx}^{*}\Big(\theta_{bl}^{tx}, \phi_{bl}^{tx}\Big)
\end{equation}

where $g_{bl}(t)$ is the complex small-scale fading gain on the $\ell$th subpath from the $b$th gNB and $\mathbf{u}_{rx}(\cdot) \in \mathbb{C}^{n_{rx}}$ and $\mathbf{u}_{tx}(\cdot) \in \mathbb{C}^{n_{tx}}$ are the precode beamforming vector for the RX and TX antenna arrays to the angular arrivals and departures.

For mmWave communication, based on the abstraction used in the prior study\cite{Yao2017}, the received power $P_{rx}$ at the receiver can be calculated as:

\begin{equation}
P_{rx} = P_{t} \cdot \psi \cdot \delta^{-1} \cdot PL^{-1}
\label{eqn:rcvdpw}
\end{equation}

where $P_{tx}$ is a reference power or transmitted power, $\psi$ is the combined antenna gain of transmitter and receiver, $\delta$ is the subpath attenuation, and $PL^{-1}$ denotes the associated line-of-sight (LOS) path-loss in dB and can be derived as: 

\begin{equation}
PL(d)~[dB]= \alpha + \beta \cdot 10log_{10}(d) + \eta
\label{eqn:pathloss}
\end{equation}

where $PL(d)$ is the mean ppath loss over a reference Tx-Rx separation distance $d$, in dB, $\alpha$ is the floating intercept in dB, $\beta$ is the ppath lossexponent, $\eta$ $\sim$ $N(0,\sigma^{2})$.


%
%

\section{Model and Problem Formulation}\label{sec:model}
In this section, first, we aim to reduce communication overhead by proving an uncoordinated network connection with no handover interruption time. Second, provide satisfactorily received signal strengths to all UEs to ensure a reliable connection. Unlike CoMP which dynamically enables the coordination of Tx/Rx over different BSs, we enable an uncoordinated transmission scheme by extending the SDN capabilities. In mmWave networks, during UE mobility, the beam realignment between transmitter and receiver is the main bottleneck. We track the UE movement and the channel gain association between UE and gNBs. The serving set of gNBs are continuously updated via the Soft-gNB within its cell. The global information of the network can be shared via the interconnection of Soft-gNBs through backhaul network as shown in Figure~\ref{fig:network}.

\subsection{Virtualized Mobility Management (vMM)}
\begin{figure}[htbp]
\centerline{\includegraphics[width=0.95\columnwidth]{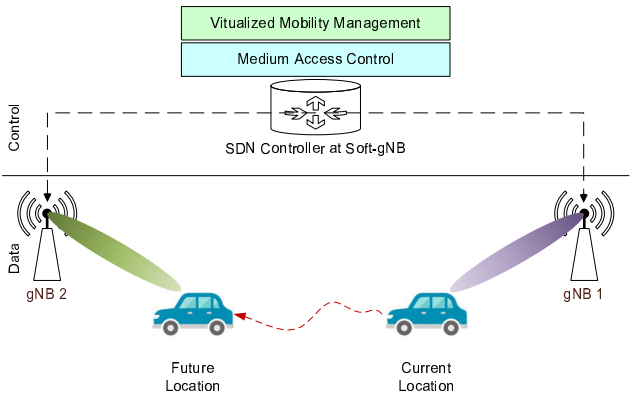}}
\caption{Overview of functional capabilities of Soft-gNB.}
\label{fig:sdnover}
\end{figure}

To achieve this, specifically, we utilize a virtualized mobility management (vMM) layer, inspired by \cite{Prados-Garzon2017}, as shown in Figure~\ref{fig:sdnover}. Each Soft-gNB in the network is enabled with vMM functionality. The vMM is in charge of processing and producing the mobility state of the UE. Specifically, we virtualized both the gNBs and UEs at the vMM using the context information and subsequently estimate the associations between gNBs and UEs. Based on these estimations, the Soft-gNB generate the beamforming weights for serving/target gNB and UE for changing the anchor point during a handover. This approach manages mobility in a scalable fashion while optimally utilizing the available resources. This implies that the disruption in channel gain during mobility can be completely solved, particularly when SDN is placed as an architectural enabler and channel strength can be estimated by the vMM. Figure~\ref{fig:sdnover} illustrates the overview of the capabilities and functionality of the proposed architecture.

To further increase the scalability of Soft-gNB and coordination among serving gNBs, we jointly shared the gNB-UE association information with gNBs involved in the serving cluster, which further limits the control message forwarding delay among the serving gNBs. As illustrated in Figure~\ref{fig:sdnover}, the gNB-1 is the serving gNB which is a member of the serving cluster; and gNB-2 is the target gNB outside the serving cluster which will be added into the serving cluster. Thus, the members of the serving-cluster need to be updated during UE moves from current location to future location. The serving-cluster updation is directly handled by the Soft-gNB by engaging vMM without the control information exchange or coordination among the gNBs. This uncoordinated handover mechanism helps the system to reduce service interruption time to meet latency requirement. In Section~\ref{subsec:cluster}, the serving-cluster formation and updation is further discussed.

\subsection{Serving Cluster Formation}\label{subsec:cluster}
In mmWave networks, it is important to guard against the channel gain loss during mobility. Thus, we propose a group of gNBs may form a cluster which can ensure that there is at least one serving-gNB connected to the UE at all times. Furthermore, this will ensure that the UE always has a connection within the network when switching to another available radio link without interrupting service. It should also ensure that no packets are lost or delayed during handover. The network may also transmit through multiple radio links across multiple gNBs. By doing so, we fulfilled all three requirements to support reliable transmission in high mobility scenarios such as seamless handover, availability of redundant link, and a robust transmission scheme.
\par
To achieve seamless handover to the target gNB, first, we use the context information and channel status of UE-gNB. The Soft-gNB receive the UEs signaling via serving-gNB and forward to the vMM, which imply an exchange of signaling messages between the vMM and UEs. When the vMM receives UE information, it processes it and later shared UE-gNB estimation with other gNB via Soft-gNB. This simplifies the control information exchange between the target gNB and UE. Second, we established the path between the target gNB and core network for downlink data transmission before the UE establishes the connection to the gNB. To summarize, the cluster formation, serving-gNB, and target-gNB selection are solely determined by the Soft-gNB while taking context information and gNB-UE channel association into consideration. Thus, the proposed framework significantly minimizes the control signaling overhead during handover.

\subsection{Service Requirements}\label{subsec:service}
In mmWave communication, the spatial fading results from the movement by the receiver or transmitter and closely linked to temporal fading. Spatial fading is experienced as variations in the received power at the receiver as it moves over short distances. Based on the mmWave channel models presented in Section~\ref{subsec:chmod}, we defined the service requirements of UEs on the Signal-to-Interference-Noise Ratio (SINR) constraint.  Let $\gamma_{u}$ and $\gamma_{u}^{min}$ denote the received SINR and the minimum SINR requirement of the $u \in \mathcal{U}$. Following \cite{Akdeniz2014}, the required SINR of UE can be formulated as below
\begin{equation}
\gamma_{u} = \frac{P_{rx}}{\varrho_{b}^{max} + \eta_b} \geq \gamma_{u}^{min}
\label{eq:outage}
\end{equation}

where $P_{rx}$ is the received power (\ref{eqn:rcvdpw}) from gNB $b \in \mathcal{B}$ to UE $u \in \mathcal{U}$ assuming the path loss $PL$ (\ref{eqn:pathloss}) and optimal long-term beamforming gain; $\varrho_{b}^{max}$ is the maximum average interference and $\eta_b$ is the thermal noise. 

So far, we have discussed the problem of maintaining a reliable connection is a potential challenge for mmWave communications. Additionally, we have also defined the service requirements for reliable connectivity during user mobility. In next section, we describe how each of the three components of our solution (i.e., SDN-enabled architecture, multi-gNB transmission (as shown in Fig.~\ref{fig:network}) and the moving cluster formation policy) are engineered to yield the best performance for the final user.

\section{Performance Evaluation}\label{sec:evaluation}
In this section, we first provide some details on the system model used for the performance evaluation and then demonstrate the performance of the multi-gNB mmWave transmission scheme through simulations. 

\subsection{System Model}
The mmWave module enables simulations for UE packet-loss with a realistic channel model that is based on 3GPP specification. The basic system parameters used for simulation are listed in Table~\ref{tbl:sys_par}. The related parameters are based on \cite{Sahoo2017, Yao2017}. An illustration of the network model is shown in Fig.~\ref{fig:network}. The network scenario consists of three gNBs uniformly distributed in a given area. Each gNB's cell with the radius equal to 100 m and users perform downlink operation over NLoS links. UE is moving through the edge of multiple cells with varying speed (30$-$ 90km/h) and gNBs are at fixed positions; and Soft-gNBs is located in the center of the cell. The UE can receive downlink transmission from multiple gNBs simultaneously. Furthermore, we assume, the simultaneous reception can be achieved by proper beamforming at UE side to mitigate interference, which is beyond the scope of this paper.

\begin{table}[htbp]
\centering
\caption{Simulation Parameters}
\label{tbl:sys_par}
\begin{IEEEeqnarraybox}[\IEEEeqnarraystrutmode\IEEEeqnarraystrutsizeadd{2pt}{1pt}]{v/c/v/c/v}
\IEEEeqnarrayrulerow\\&\mbox{\bf Parameter}&&\mbox{\bf Value}&\\\IEEEeqnarraydblrulerow\\
\IEEEeqnarrayseprow[3pt]\\&\mbox{Carrier frequency}&&$28 GHz$\hfill&\IEEEeqnarraystrutsize{0pt}{0pt}\\
\IEEEeqnarrayseprow[3pt]\\\IEEEeqnarrayrulerow\\\IEEEeqnarrayseprow[3pt]\\&\mbox{System bandwidth}&&$1 GHz$\hfill&\IEEEeqnarraystrutsize{0pt}{0pt}\\
\IEEEeqnarrayseprow[3pt]\\\IEEEeqnarrayrulerow\\\IEEEeqnarrayseprow[3pt]\\&\mbox{Transmit power ($P_{t}$)}&&$37 dBm$\hfill&\IEEEeqnarraystrutsize{0pt}{0pt}\\
\IEEEeqnarrayseprow[3pt]\\\IEEEeqnarrayrulerow\\\IEEEeqnarrayseprow[3pt]\\&\mbox{Shadow fading ($\sigma$)}&&$8.2 dB$\hfill&\IEEEeqnarraystrutsize{0pt}{0pt}\\
\IEEEeqnarrayseprow[3pt]\\\IEEEeqnarrayrulerow\\\IEEEeqnarrayseprow[3pt]\\&\mbox{Background noise (AWGN)}&&$-174 dBm/Hz$\hfill&\IEEEeqnarraystrutsize{0pt}{0pt}\\
\IEEEeqnarrayseprow[3pt]\\\IEEEeqnarrayrulerow\\\IEEEeqnarrayseprow[3pt]\\&\mbox{Minimum SINR ($\gamma_{u}^{min}$)}&&$-10 dB$\hfill&\IEEEeqnarraystrutsize{0pt}{0pt}\\
\IEEEeqnarrayseprow[3pt]\\\IEEEeqnarrayrulerow\\\IEEEeqnarrayseprow[3pt]\\&\mbox{Number of users ($N$)}&&$100 per cell$\hfill&\IEEEeqnarraystrutsize{0pt}{0pt}\\
\IEEEeqnarrayseprow[3pt]\\\IEEEeqnarrayrulerow
\end{IEEEeqnarraybox}
\end{table}

\subsection{Results and Discussion}
The performance metric considered for the evaluation obtained from multiple independent runs. The average success rate measures the packet-loss ratio of the received packet with respect to the originally transmitted packets.

\begin{figure}[htbp]
\centerline{\includegraphics[width=0.9\columnwidth]{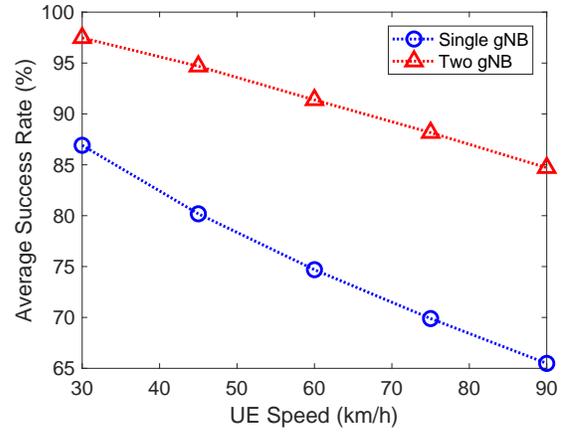}}
\caption{The average success rate over mmWave links on user mobility.}
\label{fig:result1}
\end{figure}

Figure~\ref{fig:result1} compare the average success rate when the multi-gNB transmission is not used with a multi-gNB transmission. The simulation result shows the multi-gNB transmission framework always performs better than the single downlink link. When the multi-gNB transmission is not used the success rate significantly deteriorates with increasing user speed. On the lower speed, the multi-gNB scheme performs better. However, at peak speed, at 90 km/h, relatively not performing well. The result demonstrates the redundant transmission towards UE can improve the transmission-reception failure rate, thus reliability. Thus, we conclude that the multiple downlink transmission indeed more efficient with respect to single-packet losses in the mmWave channel. Further computing capability at the gNBs could help to improve the latency by reducing the round trip time. Additionally, if an added number of gNBs are further considered to serve the UE, the trends of this curve will improve further. The multigNB mmWave framework also produces several interesting research directions worth for further study.

\section{Conclusion}\label{sec:con}
This paper investigates reliable transmission over mmWave link during UE mobility. The communication at mmWave frequency enables potential high data rate applications. However, challenges are numerous due to the unreliable and possible outage-prone mmWave links. The future ultra-dense networks demand a high flexibility to address overwhelming communication overhead. We proposed a flexible architecture based on SDN with vMM functionality. SDN introduces additional robustness and flexibility to the network. We design an adaptive framework to minimize service interruption time and ensure reliability while supporting service requirements of UEs. The proposed solution was evaluated via realistic mmWave channel model following 3GPP specifications. The results confirm the benefit introduced by SDN and multi-gNB uncoordinated transmissions.

\section*{Acknowledgment}
This work was supported by the Ministry of Science and Technology of Taiwan under Grants MOST 103-2622-E-002-034 and MOST 105-2221-E-002-014-MY3, and sponsored by MediaTek Inc.,
Hsin-chu, Taiwan.




\end{document}